\documentclass[12pt]{article}

\usepackage[left=0.75in,right=0.75in,top=0.65in,bottom=0.65in]{geometry}
\usepackage{amsmath,amssymb,bm,mathtools}
\usepackage{array,booktabs,tabularx}
\usepackage{enumitem}
\usepackage{graphicx}
\usepackage{epstopdf}
\ifdefined
\fi
\usepackage{float}
\usepackage{placeins}
\usepackage{subcaption}
\usepackage{cite}
\usepackage[hyperfootnotes=false,hidelinks]{hyperref}
\usepackage{titling}

\pretitle{\begin{center}\LARGE\bfseries}
\posttitle{\par\end{center}\vskip 0.75em}
\preauthor{\begin{center}\large}
\postauthor{\par\end{center}\vskip 0.25em}
\predate{}
\postdate{}

\newcommand{\R}{\mathbb{R}}

\newcommand{\norm}[1]{\left\lVert #1\right\rVert}
\newcommand{\dd}{\mathrm{d}}
\newcommand{\src}{\mathrm{src}}
\newcommand{\outp}{\mathrm{out}}
\newcommand{\Cset}{\mathcal{C}}
\newcommand{\Vset}{\mathcal{V}}

\newcommand{\InterfaceBandSourceError}{\ensuremath{3.069\times10^{-2}}}
\newcommand{\InterfaceValidationError}{\ensuremath{1.46\times10^{-16}}}
\newcommand{\CoupledValidationError}{\ensuremath{2.448\times10^{-2}}}
\newcommand{\LinearPhaseRMS}{\ensuremath{1.013\times10^{-2}}}
\newcommand{\DistortedPhaseRMS}{\ensuremath{2.316\times10^{-2}}}
\newcommand{\LinearMeanFidelity}{\ensuremath{0.999986}}
\newcommand{\DistortedMeanFidelity}{\ensuremath{0.999773}}

\newcommand{\OrderFourError}{\ensuremath{2.35\times10^{-2}}}

\newcommand{\OrderTwoFidelity}{\ensuremath{0.99999981}}
\newcommand{\ParameterRecoveryError}{\ensuremath{5.27\times10^{-15}}}
\newcommand{\LinearPhaseMAE}{\ensuremath{9.82\times10^{-3}}}
\newcommand{\DistortedPhaseMAE}{\ensuremath{1.74\times10^{-2}}}
\newcommand{\WeakForceSourceError}{\ensuremath{6.19\times10^{-3}}}
\newcommand{\HighForceSourceError}{\ensuremath{1.59\times10^{-1}}}
\newcommand{\StrongForceSourceError}{\ensuremath{1.09\times10^{-1}}}
\newcommand{\WeakForcePhase}{\ensuremath{1.64\times10^{-2}}}
\newcommand{\StrongForcePhase}{\ensuremath{3.65\times10^{-2}}}
\newcommand{\TwoModeSpeedup}{\ensuremath{1.33}}
\newcommand{\FourModeSpeedup}{\ensuremath{1.05}}

\newcommand{\NoiseOneMedianError}{\ensuremath{1.33\times10^{-3}}}
\newcommand{\NoiseOneParameterError}{\ensuremath{6.90\times10^{-1}}}

\newcommand{\MismatchBaseError}{\ensuremath{2.39\times10^{-1}}}
\newcommand{\MismatchStrongError}{\ensuremath{4.22\times10^{-1}}}

\newcommand{\RefinedStrongTwoError}{\ensuremath{2.04\times10^{-1}}}
\newcommand{\RefinedStrongFourError}{\ensuremath{1.13\times10^{-1}}}

\newcommand{\PODStrongFourSource}{\ensuremath{2.05\times10^{-2}}}

\newcommand{\PODStrongFourState}{\ensuremath{1.01\times10^{-5}}}
\newcommand{\ModalStrongFourSource}{\ensuremath{8.03\times10^{-2}}}

\newcommand{\RomOrderSelectionRows}{%
0.02 & 2 \\
0.04 & 2 \\
0.08 & 3 \\
0.12 & none \\
}%
\newcommand{\RomPodRows}{%
0.04 & 2 & $2.62\times10^{-2}$ & $3.50\times10^{-2}$ & $9.80\times10^{-7}$ & $2.46\times10^{-4}$ \\
0.04 & 3 & $2.52\times10^{-3}$ & $1.42\times10^{-2}$ & $3.25\times10^{-8}$ & $7.52\times10^{-5}$ \\
0.04 & 4 & $8.61\times10^{-3}$ & $4.65\times10^{-3}$ & $1.23\times10^{-8}$ & $4.30\times10^{-7}$ \\
0.12 & 2 & $1.28\times10^{-1}$ & $1.28\times10^{-1}$ & $3.77\times10^{-5}$ & $2.44\times10^{-4}$ \\
0.12 & 3 & $8.21\times10^{-2}$ & $5.19\times10^{-2}$ & $2.73\times10^{-5}$ & $8.25\times10^{-5}$ \\
0.12 & 4 & $8.03\times10^{-2}$ & $2.05\times10^{-2}$ & $2.12\times10^{-5}$ & $1.01\times10^{-5}$ \\
}%

\title{Phase-Sensitive Validation of Reduced-Order Nonlinear Mechanical Models with Dynamic Readout}
\author{Kazi T. Mahmood\textsuperscript{1} and M. Arif Hasan\textsuperscript{2}\thanks{Corresponding author: \texttt{hasan.arif@wayne.edu}}\\[0.65em]
{\footnotesize \textsuperscript{1}Department of Mechanical and Manufacturing Engineering,}\\
{\footnotesize Bowling Green State University, Bowling Green, OH 43403}\\[0.25em]
{\footnotesize \textsuperscript{2}Department of Mechanical Engineering,}\\
{\footnotesize Wayne State University, Detroit, MI 48202}}
\date{}

\begin{document}
\maketitle

\begin{abstract}
Reduced-order models of nonlinear structures are often evaluated using mechanical response errors, while the signals of interest may also pass through dynamic readout channels that modify their amplitudes and phases. We develop an offline validation framework that distinguishes mechanical source error, readout-model prediction error, and physical distortion of a normalized two-component modal state. We couple a Galerkin reduced source to two calibrated resonant channels and evaluate modal amplitudes, relative phase, and state fidelity using consistent complex coefficients. For the linear readout model, an exact output-error decomposition distinguishes source and interface modeling contributions, while a two-channel identity relates differential channel transfer to normalized-state distortion. For the ten-mass nonlinear lattice, the two-mode source and coupled-output errors over the two-resonance interface band are $\InterfaceBandSourceError$ and $\CoupledValidationError$. At $F_0=0.04$, the relative-phase mean absolute error increases from $\LinearPhaseMAE$ rad for the linear interface to $\DistortedPhaseMAE$ rad for the nonlinear interface. At the strongest forcing, a four-coordinate proper orthogonal decomposition model gives a source error of $\PODStrongFourSource$, compared with $\ModalStrongFourSource$ for four linear modes on the same held-out test set. The noise analysis shows that accurate in-band output prediction can coexist with substantial parameter error, while a constrained linear approximation of nonlinear readout data produces a substantial residual. We use these findings to assess reduced-model adequacy for the specified outputs, forcing levels, and validation frequencies.
\end{abstract}

\section{Introduction}
\label{sec:introduction}

Nonlinear mechanical and phononic lattices support vibration and wave responses that depend on architecture, coupling, and constitutive behavior \cite{Nesterenko2001,Porter2015,PatilMatlack2022}. Predicting these responses requires resolving the response to an applied excitation over the intended operating range. At finite forcing amplitudes, nonlinear restoring forces can shift resonance peaks and couple modal responses \cite{NayfehMook1979,Kerschen2006}. The resulting vibration depends on both modal amplitude and relative phase \cite{VolvertKerschen2021}. These modal contributions combine to determine the physical displacement at an observation point. However, agreement between the reduced and full mechanical responses alone does not characterize the amplitude and phase changes introduced by a subsequent dynamic readout. We therefore evaluate mechanical source accuracy, readout prediction accuracy, and the preservation of modal relationships as distinct aspects of model validation.

Projection-based reduction represents a mechanical response within a lower-dimensional trial space \cite{Benner2015}. Linear eigenmodes are useful when a limited set of modes dominates the response, while displacement-based proper orthogonal decomposition (POD) identifies basis vectors from representative snapshots \cite{Kerschen2005}. Modal derivatives, quadratic manifolds, and invariant-manifold approaches address additional features of nonlinear response geometry \cite{ShawPierre1993,Jain2017,Rutzmoser2017,HallerPonsioen2016,Touze2021}. We distinguish these methods of constructing reduced representations from goal-oriented approaches that assess or improve selected output quantities \cite{Hoang2015,Borggaard2016,Yano2020}. In the present study, we retain conventional modal and POD bases and evaluate their source, readout, and phase-sensitive state errors under a common validation procedure.

Two selected modal channels provide a compact representation of their relative amplitude and phase, which can be examined separately from the complete mechanical displacement response. Experiments on driven granular systems have demonstrated classical analogues of coherent two-state superpositions \cite{RN7}. A normalized pair of complex modal coefficients can be represented on the Bloch sphere using an amplitude-balance angle and a relative-phase angle \cite{RN156}. Related elastic-bit studies have examined controlled state manipulation and classical analogues of logical operations \cite{RN21}. We use this established two-component representation to assess the readout of the first two modal coordinates of the nonlinear lattice. A common complex channel gain preserves the normalized state, whereas unequal channel gains or phases can alter it.

Dynamic readout elements can introduce additional amplitude and phase responses between a mechanical source and an acquired signal. Resonant transducers and filtering stages have frequency-dependent characteristics that must be considered when interpreting measurements \cite{VlajicChijioke2016,Aspelmeyer2014}. In the numerical model considered here, two independent damped resonant channels receive prescribed source modal coordinates through one-way coupling. Their unequal responses change the readout-channel amplitude ratio and relative phase. We distinguish the accuracy of predicting these outputs from the physical source-to-output distortion produced by the channels.

A common transport or acquisition delay also affects complex waveform alignment. For the two modal coefficients evaluated at the same drive frequency, the corresponding common phase factor cancels from their relative phase and normalized-state fidelity. We therefore correct the common delay while retaining channel-dependent phase shifts in the state-transfer assessment.

We aim to evaluate reduced mechanical models through three distinct comparisons: agreement with the full-order mechanical response, prediction of the dynamic readout output, and preservation of the normalized two-channel modal state. We employ Galerkin reduction, fixed modal and POD bases, and a calibrated two-channel interface to examine these comparisons over specified forcing amplitudes and frequency grids. We use an exact linear error decomposition and a channel-distortion identity to interpret how source truncation and unequal readout dynamics affect the respective metrics. We also examine noise-dependent parameter recovery and the residual produced by fitting a constrained linear interface model to nonlinear readout data. The resulting tolerance-based order comparisons use full-order reference responses and constitute offline model validation.

The remainder of the paper is organized as follows. Section~\ref{sec:framework} presents the coupled source--interface formulation, calibration procedure, and metrics for model accuracy and state preservation. Section~\ref{sec:numerical} describes the nonlinear lattice, readout interface, and numerical validation design. Section~\ref{sec:results} examines source-ROM convergence, order and basis selection, interface identification, and the effects of forcing amplitude on state transfer. Section~\ref{conclusion} summarizes the findings and directions for further validation. Supporting derivations and additional numerical checks are provided in the Supplementary Material.

\section{Source--Interface Reduced-Order Framework}
\label{sec:framework}

\begin{figure}[H]
\centering
\includegraphics[width=\linewidth,trim=4bp 4bp 4bp 4bp,clip]{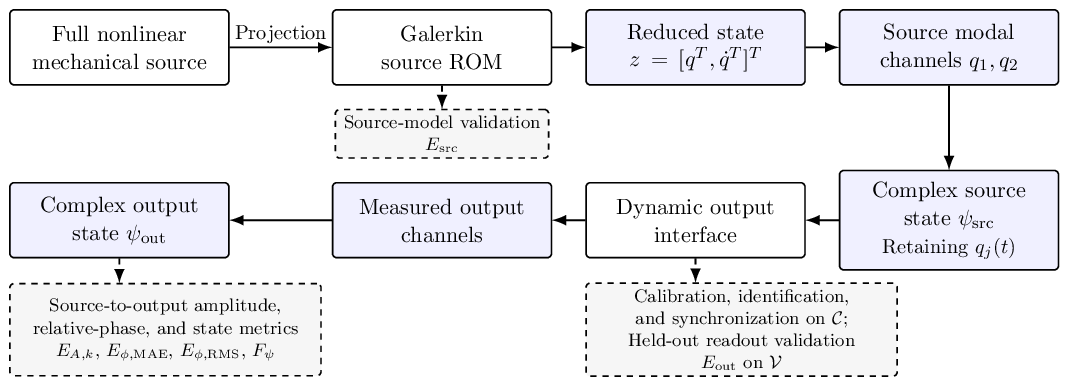}%
\caption{Source-to-output modeling and validation. Galerkin reduction defines the source ROM, and calibration determines the interface parameters. The source-state block retains the modal signals $q_j(t)$ that drive the interface; the normalized state $\psi_{\rm src}$ is used for state comparisons. The validation metrics assess source accuracy, output prediction, and physical state preservation.}
\label{fig:framework}
\end{figure}

We model the nonlinear mechanical source and dynamic readout as a one-way cascade driven by selected source modal coordinates. Galerkin projection reduces the source equations, and independently calibrated readout channels transform the modal responses into output signals. We use common complex-coefficient conventions to compare source-model error, readout prediction error, and physical changes in modal amplitude balance and relative phase. Figure~\ref{fig:framework} summarizes the computational and validation stages.

The first two source modal coordinates drive the readout channels, while their complex fundamental coefficients define the normalized source state $\psi_{\rm src}$. The corresponding calibrated output coefficients define the normalized output state $\psi_{\rm out}$. We use the source-response error $E_{\rm src}$ to compare the reduced and full mechanical source observations, and the readout prediction error $E_{\rm out}$ to compare predicted and reference readout outputs. We evaluate physical transfer distortion using the channel-amplitude error $E_{A,k}$, the relative-phase errors $E_{\phi,{\rm MAE}}$ and $E_{\phi,{\rm RMS}}$, and the normalized-state fidelity $F_\psi$. These metrics are defined in Sec.~\ref{sec:metrics} and are interpreted separately because they characterize different aspects of source accuracy, readout prediction, and state preservation.

\subsection{Nonlinear mechanical source and Galerkin reduction}
\label{sec:source_rom}

The full-order mechanical model (FOM) is,
\begin{equation}
M\ddot u+C\dot u+Ku+f_{\rm nl}(u,\dot u;\mu)=B\Gamma(t),
\label{eq:fom}
\end{equation}
where $u\in\R^N$, $M,C,K\in\R^{N\times N}$ are the mass, damping, and stiffness matrices, $f_{\rm nl}$ contains the nonlinear internal forces, and $B\Gamma(t)$ is the applied load. For a fixed basis $\Phi_r\in\R^{N\times r}$, the reduced displacement coordinates $q(t)\in\R^r$ define the approximation $u\approx\Phi_rq$. Applying the Galerkin condition $\Phi_r^T\mathcal R=0$ \cite{Benner2015} gives,
\begin{equation}
M_r\ddot q+C_r\dot q+K_rq+f_r(q,\dot q;\mu)=B_r\Gamma(t),
\label{eq:rom_second}
\end{equation}
with $M_r=\Phi_r^TM\Phi_r$, $C_r=\Phi_r^TC\Phi_r$, $K_r=\Phi_r^TK\Phi_r$, $B_r=\Phi_r^TB$, and,
\begin{equation}
f_r(q,\dot q;\mu)=\Phi_r^Tf_{\rm nl}(\Phi_rq,\Phi_r\dot q;\mu).
\label{eq:projected_force}
\end{equation}
The projected nonlinear force retains cubic interactions among the retained coordinates. The reduced dynamics approximate the influence of omitted modal coordinates through the selected trial space. We precompute the linear operators and spring-extension basis and evaluate the full projected cubic spring law without nonlinear-force hyper-reduction. Defining $z=[q^T,\dot q^T]^T\in\R^{2r}$ gives the first-order source model,
\begin{equation}
\begin{aligned}
\dot z&=A_rz+F_r(z;\mu)+B_z\Gamma(t),\\
A_r&=\begin{bmatrix}0&I\\-M_r^{-1}K_r&-M_r^{-1}C_r\end{bmatrix},\quad
F_r=\begin{bmatrix}0\\-M_r^{-1}f_r\end{bmatrix},\quad
B_z=\begin{bmatrix}0\\M_r^{-1}B_r\end{bmatrix}.
\end{aligned}
\label{eq:rom_first}
\end{equation}
We define a physical source observation by $x_{\rm src}=Pz$. A displacement selected by $L_u$ uses $P=[L_u\Phi_r\;0]$. In the numerical example, this observation is the displacement of mass 7, while the first two modal coordinates separately drive the readout channels. We therefore assess physical displacement accuracy and modal-channel accuracy as distinct reduced-model properties.

We construct the primary reduced basis from mass-normalized linear modes, for which $\Phi_r^TM\Phi_r=I$. A displacement-snapshot proper orthogonal decomposition (POD) basis provides a second fixed linear trial space at equal retained dimension \cite{Kerschen2005}. We use the same constitutive force law, projection procedure, solver settings, and physical-output definition for both bases. Modal derivatives, quadratic manifolds, and invariant-manifold approaches offer other ways to represent nonlinear response geometry \cite{Jain2017,Rutzmoser2017,HallerPonsioen2016,Touze2021}; we use the two fixed linear spaces to isolate basis dependence in the reported comparisons. The Supplementary Material gives the projection and normalization details.

\subsection{Dynamic output-interface model}
\label{sec:interface_model}

We treat the readout interface as a dynamical subsystem driven by the source through one-way coupling. A general nonlinear state-space representation describes internal readout dynamics, source excitation, deterministic nonlinear effects, and disturbances \cite{AstromMurray2008}. We write the interface states relative to an operating equilibrium and represent the common acquisition or transport delay in the source input. The resulting model is,
\begin{equation}
\begin{aligned}
\dot y(t)&=A_o(\eta)y(t)+G_oz(t-\tau_{\rm sync})
 +D_o(y,z;\eta)+w(t),\\
x_{\rm out}(t)&=Qy(t)+v(t).
\end{aligned}
\label{eq:interface_general}
\end{equation}
where $\eta$ contains interface parameters, $w$ and $v$ denote process and measurement disturbances, and $\tau_{\rm sync}$ is a common transport or acquisition delay. Parameters fitted to calibration data define an empirical interface model whose predictions are evaluated on a disjoint validation set.

The interface state $y$ represents internal readout coordinates rather than the mechanical displacement of the source. Depending on the physical realization, such coordinates may describe resonator or filter states. The matrices $A_o(\eta)$ and $G_o$ specify internal linear dynamics and the map from delayed source coordinates to interface excitation, while $Q$ maps interface states to output signals. The numerical realization below specializes this general description to two independently driven resonant channels.

In Eq.~\eqref{eq:interface_general}, $D_o$ collects deterministic effects beyond the nominal linear interface dynamics and uses the delayed source state as its source argument. The process disturbance $w$ represents unresolved interface forcing, and $v$ represents measurement noise. For the numerical realization, we retain cubic restoring nonlinearities within each channel, without cross-channel coupling. The synthetic identification noise is introduced in complex frequency-domain output data rather than through a time-domain simulation of $w$ or $v$. Further formulation details appear in the Supplementary Material.

For the numerical study, we prescribe the first two simulated modal coordinates as inputs to two independent resonant transducer channels. This choice isolates their dynamic gain and phase effects; it does not model the additional inverse problem of recovering modal coordinates from sparse physical sensors. Each channel follows a damped second-order equation with a possible cubic restoring term,
\begin{equation}
\ddot y_j+2\zeta_{o,j}\omega_{o,j}\dot y_j+\omega_{o,j}^2y_j
+\chi_jy_j^3=g_jq_j(t-\tau_{\rm sync}),\qquad j=1,2.
\label{eq:interface_specific}
\end{equation}
Setting $\chi_j=0$ gives the linear interface used for parameter identification. Positive $\chi_j$ produces a hardening cubic contribution whose influence depends on output amplitude. We use the nonlinear interface to study transfer distortion and the residual obtained when a constrained linear readout model is fitted to nonlinear data.

For each channel, $\omega_{o,j}$ and $\zeta_{o,j}$ specify resonance frequency and damping, $g_j$ determines coupling gain, and $\chi_j>0$ introduces a hardening restoring term \cite{NayfehMook1979}. The two output signals provide separate complex coefficients for state reconstruction. Reconstruction from a single physical measurement would require an additional observation model and is not part of the numerical realization. For $\chi_j=0$, the transfer function at angular frequency $\Omega$ is,
\begin{equation}
H_j(\Omega)=
\frac{g_j e^{-i\Omega\tau_{\rm sync}}}
{\omega_{o,j}^2-\Omega^2+i2\zeta_{o,j}\omega_{o,j}\Omega}.
\label{eq:interface_transfer}
\end{equation}
The magnitude of $H_j$ determines the channel amplitude response, and its argument includes the common transport phase and the channel's dynamic phase. Differences in $\omega_{o,j}$ or $\zeta_{o,j}$ produce differential gain and phase, which alter the normalized output state. A nonzero cubic coefficient adds amplitude dependence to the effective stiffness and phase response.

\subsection{Calibration, synchronization, and identification}
\label{sec:calibration}

In Eq.~\eqref{eq:interface_transfer}, the total linear transfer phase consists of the common transport term $-\Omega\tau_{\rm sync}$ and the channel-dependent term $\phi_{H,j}(\Omega)$. We estimate the common delay jointly with the interface parameters and retain the differential dynamic phase when comparing source and output states. For components observed at the same drive frequency, common-delay correction changes waveform alignment but not the two-channel relative phase.

Let $\Cset$ and $\Vset$ denote disjoint calibration and validation frequency sets. Interface parameters are estimated only on $\Cset$,
\begin{equation}
\hat\eta=\arg\min_{\eta}\sum_{\Omega_k\in\Cset}
\norm{S^{-1}\left[\hat x_{\rm out}(\Omega_k)-
\hat x_{\rm out}^{\rm model}(\Omega_k;\eta)\right]}_2^2,
\label{eq:id_cost}
\end{equation}
where the diagonal entries of $S$ are the median calibration magnitudes of the output channels. This scaling balances the channel contributions to the complex least-squares residual. Identified parameters are fixed before evaluating prediction errors on $\Vset$ \cite{Ljung1999,NoelKerschen2017}. Cross-correlation can provide an initial estimate of $\tau_{\rm sync}$ for the joint fit \cite{KnappCarter1976}.

We identify the seven-parameter vector $\eta=[\omega_{o,1},\omega_{o,2},\zeta_{o,1},\zeta_{o,2},g_1,g_2,\tau_{\rm sync}]^T$ from the selected calibration band. The fitted response may be accurately predicted within that band even when individual parameters are weakly constrained. Parameter bounds and initialization are therefore part of the identification protocol and must be considered when interpreting recovered physical values.

We use the noiseless matched-model case as an implementation check and the perturbed-data cases to examine output prediction under parameter uncertainty and model mismatch. For comparison of physical source and output states, we normalize the channels by their linear static gains and remove the common delay using,
\begin{equation}
c_{{\rm out},j}(\Omega)=
\frac{\omega_{o,j}^2}{g_j}\,\hat y_j(\Omega)
e^{i\Omega\tau_{\rm sync}}.
\label{eq:calibrated_output_channel}
\end{equation}
The factor $\omega_{o,j}^2/g_j$ normalizes the linear zero-frequency gain. Equation~\eqref{eq:calibrated_output_channel} also removes the specified common delay, but it retains the frequency-dependent magnitude and phase of $\omega_{o,j}^2/(\omega_{o,j}^2-\Omega^2+i2\zeta_{o,j}\omega_{o,j}\Omega)$. The resulting coefficients therefore describe the physical dynamic transfer remaining after static-gain normalization, rather than full inversion of the interface.

For the fundamental coefficients of both channels at the same $\Omega$, a common delay multiplies the two coefficients by the same unit-modulus factor. It therefore cancels in their relative phase and normalized-state fidelity. This invariance concerns the two-component state comparison; output waveforms and complex prediction residuals still require consistent timing. Unequal channel transfer phases remain in the physical distortion metrics.

\subsection{Complex source and output states}
\label{sec:complex_states}

At each drive frequency $\Omega$, the complex coefficients $c_1(\Omega)$ and $c_2(\Omega)$ specify the amplitudes and phases of the two channels.

The coefficients weight the selected vibration modes, defined by $K\phi_j=\omega_j^2M\phi_j$, where $\phi_j$ is the $j$th mode shape and $\omega_j$ is its natural angular frequency. The first two mode shapes are denoted by $|E_1\rangle=\phi_1$ and $|E_2\rangle=\phi_2$, with mass normalization $\phi_j^TM\phi_j=1$ and orthogonality $\phi_1^TM\phi_2=0$.
For a symmetric two-coordinate system with unit masses, the in-phase and out-of-phase modes are \cite{RN156},
\[
|E_1\rangle_{(2)}=\frac{1}{\sqrt{2}}
\begin{bmatrix}1\\1\end{bmatrix},
\qquad
|E_2\rangle_{(2)}=\frac{1}{\sqrt{2}}
\begin{bmatrix}1\\-1\end{bmatrix}.
\]
The numerical study uses the first two eigenmodes of the fixed--fixed ten-mass chain. For $M=I$ and identical linear springs, their entries are $(\phi_j)_n=\sqrt{2/11}\sin(jn\pi/11)$ for $n=1,\ldots,10$ and $j=1,2$, up to a consistent eigenvector sign.

Normalizing the coefficient vector defines the two-component state
\begin{equation}
\psi(\Omega)=\begin{bmatrix}\alpha(\Omega)\\\beta(\Omega)\end{bmatrix}
\equiv\frac{1}{\sqrt{|c_1(\Omega)|^2+|c_2(\Omega)|^2}}
\begin{bmatrix}c_1(\Omega)\\c_2(\Omega)\end{bmatrix},
\qquad \norm{\psi}_2=1.
\label{eq:state_definition}
\end{equation}
For $|c_1|^2+|c_2|^2>0$, the normalized coefficients preserve the ratio of modal amplitudes and their relative phase while discarding common magnitude. This state characterizes the selected pair of modes rather than the entire ten-coordinate mechanical displacement. In the eigenmode basis,
\[
|\psi(\Omega)\rangle
=\alpha(\Omega)|E_1\rangle+\beta(\Omega)|E_2\rangle.
\]
Factoring out the common phase of $c_1$ gives the Bloch-state representation
\begin{equation}
\begin{aligned}
\psi(\Omega)&\sim
\begin{bmatrix}
\cos\!\left[\theta(\Omega)/2\right]\\
e^{i\varphi(\Omega)}\sin\!\left[\theta(\Omega)/2\right]
\end{bmatrix},\\
\theta(\Omega)&=2\operatorname{tan^{-1}}\!\left(|\beta(\Omega)|/|\alpha(\Omega)|\right),\\
\varphi(\Omega)&=\arg\!\left[\beta(\Omega)\right]-\arg\!\left[\alpha(\Omega)\right]
=\arg\!\left[\beta(\Omega)\alpha^*(\Omega)\right].
\end{aligned}
\label{eq:global_phase_removed}
\end{equation}
The polar angle $\theta\in[0,\pi]$ describes the channel amplitude partition, and the azimuthal angle $\varphi$ gives the phase of channel 2 relative to channel 1. The polar angle is defined at both poles. Relative-phase statistics require both channel amplitudes to be resolved because $\varphi$ is undefined when either coefficient vanishes. The factor $e^{+i\varphi}$ in the second component fixes the phase-sign convention.

For the source state, $c_{{\rm src},j}=\hat q_j$, where $\hat q_j(\Omega)$ is the complex fundamental Fourier coefficient of the $j$th selected modal coordinate. The output coefficients $c_{{\rm out},j}$ follow Eq.~\eqref{eq:calibrated_output_channel}. Each pair is normalized using Eq.~\eqref{eq:state_definition} and mapped to the angles in Eq.~\eqref{eq:global_phase_removed}.

For measurements of physical coordinates, the selected modal coefficients can be estimated from a sampled modal basis by least squares \cite{Tarpo2020}. Such reconstruction depends on sufficient sensor locations and the conditioning of the sampled mode-shape matrix. The present numerical evaluation instead uses complete simulated coordinates and mass-normalized projection onto the first two linear modes. The Supplementary Material provides the reconstruction formula and related two-channel constructions \cite{RN21,RN156}.

The steady-state coefficients define an operating-point path $\psi(\Omega)$ as the drive frequency is varied; this path is not a time-dependent state trajectory. Time-resolved quantities such as phase drift or geometric-phase accumulation would require complex envelopes obtained by lock-in demodulation, local Fourier analysis, or an analytic-signal method \cite{Feldman2011}. We use only fundamental steady-state coefficients in the reported frequency-sweep state metrics, with phase comparisons restricted to resolved modal components.

\subsection{Validation and state-preservation metrics}
\label{sec:metrics}

At a fixed forcing amplitude, let $H_N(\Omega)$ and $H_r(\Omega)$ denote the complex fundamental displacement responses at the selected physical observation point for the full-order and reduced source models over $[\Omega_a,\Omega_b]$. Their relative band error is
\begin{equation}
E_{\rm src}=\left[
\frac{\int_{\Omega_a}^{\Omega_b}|H_r(\Omega)-H_N(\Omega)|^2\,\dd\Omega}
{\int_{\Omega_a}^{\Omega_b}|H_N(\Omega)|^2\,\dd\Omega}
\right]^{1/2},
\label{eq:Esrc}
\end{equation}
This band norm combines magnitude and phase discrepancies in the physical source observation. For the separate phase comparison, $\Omega_k$ denotes a sampled drive frequency and $\mathcal M$ contains indices for which $|H_N(\Omega_k)|$ exceeds $5\%$ of its maximum over the band. We compute the RMS phase discrepancy over $n_m=|\mathcal M|$ retained samples using
\begin{equation}
e_{\phi,{\rm src},k}=\!\left[\arg H_r(\Omega_k)-\arg H_N(\Omega_k)\right],
\qquad
E_{\phi,{\rm src,RMS}}=\left(\frac{1}{n_m}\sum_{k\in\mathcal M}e_{\phi,{\rm src},k}^2\right)^{1/2}.
\label{eq:source_phase_rms}
\end{equation}

For output validation, $\hat x_{\rm out}^{\rm model}$ and $\hat x_{\rm out}^{\rm true}$ denote the predicted and reference complex output coefficients. Their relative error on the held-out frequency set $\Vset$ is,
\begin{equation}
E_{\rm out}=\frac{\norm{\hat x_{\rm out}^{\rm model}-\hat x_{\rm out}^{\rm true}}_{2,\Vset}}
{\norm{\hat x_{\rm out}^{\rm true}}_{2,\Vset}}.
\label{eq:Eout}
\end{equation}
The norm $\norm{\cdot}_{2,\Vset}$ is the Euclidean norm of the output coefficients stacked over all channels and validation frequencies.

The interface is driven by the first two physical modal channels introduced in Sec.~\ref{sec:complex_states}. Their complex fundamental Fourier coefficients are collected in $\hat q_\ell^{(2)}(\Omega)=[\hat q_{\ell,1}(\Omega),\hat q_{\ell,2}(\Omega)]^T$, where $\ell=N$ denotes the full-order response and $\ell=r$ the reconstructed reduced response. Both vectors use the same mass-normalized physical mode shapes; the superscript $(2)$ denotes the two selected channels. Their relative error is,
\[
E_q=\frac{\norm{\hat q_r^{(2)}-\hat q_N^{(2)}}_2}{\norm{\hat q_N^{(2)}}_2},
\]
with coefficients stacked over both channels and the sampled drive frequencies.

At each frequency, let $H=\operatorname{diag}(H_1,H_2)$ be the reference interface transfer matrix from Eq.~\eqref{eq:interface_transfer}, and let $\widehat H$ be its identified approximation. The reference output coefficients are $\hat y_N=H\hat q_N^{(2)}$, and the prediction driven by the reduced source is $\hat y_r=\widehat H\hat q_r^{(2)}$. Adding and subtracting $\widehat H\hat q_N^{(2)}$ gives,
\begin{equation}
\hat y_r-\hat y_N
=\widehat H(\hat q_r^{(2)}-\hat q_N^{(2)})
+(\widehat H-H)\hat q_N^{(2)}.
\label{eq:error_decomposition}
\end{equation}
The first term represents reduced-source modal error propagated through the identified interface, and the second represents interface-model discrepancy under full-order modal excitation. The identity is exact for the linear transfer representation at each frequency. Because the contributions are complex, they can reinforce or partly cancel; a small total output residual does not by itself imply that both terms are individually small. We evaluate nonlinear-interface predictions directly rather than applying the linear decomposition.

Physical transfer from source to output is evaluated using the two-component coefficient vectors $c_s(\Omega)=[c_{s,1}(\Omega),c_{s,2}(\Omega)]^T$, with $s\in\{\src,\outp\}$. The source entries are the modal Fourier coefficients $c_{{\rm src},j}=\hat q_j$, and the output entries are calibrated according to Eq.~\eqref{eq:calibrated_output_channel}. At each frequency $\Omega_k$, taking componentwise magnitudes gives the channel-amplitude vectors $|c_{\rm src}(\Omega_k)|$ and $|c_{\rm out}(\Omega_k)|$. Their relative difference is,
\begin{equation}
E_{A,k}=\frac{\norm{|c_{\rm out}(\Omega_k)|-|c_{\rm src}(\Omega_k)|}_2}
{\norm{|c_{\rm src}(\Omega_k)|}_2}.
\label{eq:amplitude_transfer_error}
\end{equation}
Here, the metric measures amplification, attenuation, and differential gain remaining after static calibration.

Let $\Delta\phi_{s,k}$ denote the phase of channel 2 relative to channel 1 at $\Omega_k$, where the superscript $*$ denotes complex conjugation. The relative phase and its source-to-output residual are,
\begin{equation}
\begin{aligned}
\Delta\phi_{s,k}&=\arg\!\left[c_{s,2}(\Omega_k)c_{s,1}^*(\Omega_k)\right],
\\
e_{\phi,k}&=\Delta\phi_{{\rm out},k}-\Delta\phi_{{\rm src},k}.
\end{aligned}
\label{eq:phase_residual}
\end{equation}
Thus, $\Delta\phi_{s,k}=\varphi_s(\Omega_k)$ is the Bloch azimuth defined in Eq.~\eqref{eq:global_phase_removed}, and $e_{\phi,k}$ measures its change through the interface. For $n$ retained frequencies at which both channels have resolved amplitudes in the source and output, the mean absolute error (MAE) and RMS error are,
\begin{equation}
E_{\phi,{\rm MAE}}=\frac{1}{n}\sum_{k=1}^n|e_{\phi,k}|,
\qquad
E_{\phi,{\rm RMS}}=\left(\frac{1}{n}\sum_{k=1}^ne_{\phi,k}^2\right)^{1/2}.
\label{eq:phase_metrics}
\end{equation}
The MAE gives the mean absolute source-to-output phase discrepancy, and the RMS gives greater weight to larger discrepancies. Both quantify physical transfer across the sampled drive frequencies. For time-resolved data, phase drift can be estimated from the least-squares slope of the unwrapped, amplitude-masked phase residual.

Adopting the pure-state overlap form of state fidelity \cite{Jozsa1994} as a mathematical comparison metric for classical two-component modal states, the squared overlap at each frequency defines the fidelity $F_\psi$ and its complementary error $E_\psi$:
\begin{equation}
F_\psi=|\psi_{\rm src}^{H}\psi_{\rm out}|^2,
\qquad E_\psi=1-F_\psi.
\label{eq:state_fidelity}
\end{equation}
The superscript $H$ denotes the conjugate transpose. Fidelity lies in $[0,1]$ and compares normalized two-component complex states independently of common amplitude and phase. Unit fidelity corresponds to states proportional by a nonzero common complex factor. Because normalization suppresses common gain and gives less weight to phase errors in weak modal components, we interpret $F_\psi$ together with $E_A$ and the relative-phase metrics rather than as a complete measure of waveform accuracy.

\subsection{Channel distortion and output-based order selection}
\label{sec:order_selection}

For the calibrated linear interface, $t_j(\Omega)$ denotes the complex transfer factor of channel $j$ after static gain calibration and common-delay correction, as described below Eq.~\eqref{eq:calibrated_output_channel}. At a fixed frequency, $c_{\rm out}=Tc_{\rm src}$ with $T=\operatorname{diag}(t_1,t_2)$. The normalized source weight in channel 1 is $p=|\alpha_{\rm src}|^2$, and that in channel 2 is $1-p=|\beta_{\rm src}|^2$. Substituting these channel weights and transfer factors into Eq.~\eqref{eq:state_fidelity} gives,
\begin{equation}
1-F_\psi=
\frac{p(1-p)|t_1-t_2|^2}{p|t_1|^2+(1-p)|t_2|^2},
\label{eq:channel_distortion}
\end{equation}
for a nonzero output norm. A common complex transfer leaves the normalized state unchanged, whereas unequal channel factors generally change the relative amplitude or phase when both source components contribute. In states dominated by one component, a substantial error in the weaker channel can have only a small effect on fidelity. We use this algebraic relationship to interpret the physical transfer response, together with the separately defined amplitude and phase errors.

We select the smallest tested $r\geq2$ whose full-order-referenced validation errors satisfy the prescribed source, output, and modal-state tolerances. The numerical study uses illustrative limits $E_{\rm src}\leq0.05$, $E_{\rm out}\leq0.05$, and $\max_\Omega(1-F_\psi^{\rm ROM})\leq10^{-3}$, with $F_\psi^{\rm ROM}$ comparing reduced and full source states before readout. The maximum is evaluated on the stated validation frequency grid. These comparisons constitute offline acceptance tests for the sampled operating regime; they are not an online error certificate. We evaluate physical readout distortion separately from the truncation-dependent acceptance metrics.

\section{Numerical Model and Validation Design}
\label{sec:numerical}

\subsection{Nonlinear source model}

The test system is a fixed--fixed chain of ten unit masses connected by identical springs with linear and cubic restoring components. Let $D\in\mathbb{R}^{11\times10}$ be the signed spring-extension operator, including the two boundary springs, so that $Du$ contains the extension of every spring and $u_0=u_{11}=0$. The governing equation is
\begin{equation}
	M\ddot u+C\dot u+K u+\gamma D^T(Du)^{\circ 3}
	=b_3F_0\sin(\Omega t),
	\label{eq:lattice_model}
\end{equation}
where $(\cdot)^{\circ3}$ denotes elementwise cubing, $M=I$, $K$ is the linear stiffness matrix assembled from unit-stiffness springs, and $b_3$ is the force-location vector corresponding to mass 3. Each spring obeys $F_s(\Delta u)=\Delta u+\gamma(\Delta u)^3$ with $\gamma=0.5$. Rayleigh damping is $C=a_RM+b_RK$ with $a_R=b_R=0.01$. The cubic hardening term couples the linear modes through the projected nonlinear force.

We apply the external forcing to mass 3 and use the displacement of mass 7 as the physical observation for source-model accuracy. The distinct forcing and observation locations allow the observed response to reflect contributions from modes excited by the applied force. For the interface and forcing studies, we use $0.20\leq\Omega\leq0.66$ rad/s, containing the first two linear natural frequencies, $0.2846$ and $0.5635$ rad/s. The model-order study extends the range to $1.20$ rad/s to include the first four linear natural frequencies. Nonlinear resonance locations may vary with forcing amplitude, so these frequencies specify the underlying linear reference modes.

\subsection{Reduced-order models}

Mass-normalized eigenvectors of $(K,M)$ form $\Phi_r$ and are retained in ascending natural-frequency order. The reduced cubic force $\gamma\Phi_r^TD^T(D\Phi_rq)^{\circ3}$ uses the precomputed spring-extension basis $D\Phi_r$. This evaluation retains the constitutive spring law and all cubic interactions generated within the retained trial space; it does not hyper-reduce nonlinear-force evaluation. The omission of higher modal coordinates remains an approximation.

We compare orders $r=1,2,3,4$ at $F_0=0.04$ over the four-resonance band and evaluate the tested modal orders at all four forcing levels over the two-resonance band. We use the first two modal coefficients for the normalized-state comparison, which therefore requires $r\geq2$. For the separate POD comparison, we construct two-, three-, and four-coordinate models and evaluate them at forcing and frequency conditions excluded from POD training.

\subsection{Output-interface cases}

We use the two-channel interface in Eq.~\eqref{eq:interface_specific} with transducer natural frequencies of $3.0$ and $2.6$ rad/s, above the mechanical excitation band. The specified damping and gains produce different frequency-dependent amplitude and phase responses in the two channels even though both operate below their resonances. These parameters define a controlled illustrative readout configuration rather than a fit to a particular physical device.

The linear case sets $\chi_j=0$; the nonlinear case uses $\chi=[40,30]$ to represent cubic hardening in the independently driven channels. We compute nonlinear outputs by time integration using the mean and first 15 Fourier harmonics of each converged full-order source modal signal. A 31-harmonic check at representative operating points is reported in the Supplementary Material. We evaluate normalized states from the fundamental output coefficients and use the exact reference static gains and common delay for physical-transfer plots. The numerical interface is a one-way cascade, and these curves quantify the specified transfer behavior rather than uncertainty in experimentally estimated calibration factors.

\subsection{Calibration and validation design}

We partition the selected frequencies into interleaved calibration and validation sets spanning the excitation band. This protocol assesses held-out in-band interpolation rather than extrapolation to an unseen frequency region. We identify seven interface parameters, $\omega_{o,1}$, $\omega_{o,2}$, $\zeta_{o,1}$, $\zeta_{o,2}$, $g_1$, $g_2$, and $\tau_{\rm sync}$, using the bounded complex least-squares procedure in Section~\ref{sec:calibration}. The reference full-order modal coefficients drive the interface during fitting, so the identified-interface error is evaluated separately from mechanical ROM error.

The identified values are fixed before validation. The interface-only test uses full-order modal coefficients to evaluate interface prediction. The coupled test uses two-mode ROM coefficients to evaluate the combined source and interface error. Matched noiseless data verify the identification procedure.

\subsection{Response calculation}

Steady responses of the full and reduced systems are computed independently at each frequency by adaptive time integration. After transient decay, the complex fundamental coefficient of a response $x(t)$ is evaluated as,
\begin{equation}
\hat x(\Omega)=\frac{2}{T_m}\int_{t_0}^{t_0+T_m}x(t)e^{-i\Omega t}\,dt,
\label{eq:fourier_extraction}
\end{equation}
where $T_m$ denotes the steady-state observation window spanning an integer number of drive periods. We use the same fundamental-coefficient convention for physical, modal, full-order, and reduced responses. Each frequency is independently initialized from rest. After at least 36 periods, convergence requires relative changes below $10^{-6}$ in successive eight-period Fourier windows and consecutive-period displacement waveforms for two successive windows. All reported source responses meet this stated periodic convergence criterion under the specified initialization protocol. The procedure does not characterize other possible steady-state branches. The Supplementary Material documents solver, window-length, harmonic, and frequency-resolution checks. Table~\ref{tab:parameters} lists the model parameters.

\begin{table}[H]
\centering
\caption{Principal model parameters}
\label{tab:parameters}
\small
\begin{tabularx}{\textwidth}{@{}>{\raggedright\arraybackslash}p{0.20\textwidth}>{\raggedright\arraybackslash}p{0.36\textwidth}>{\raggedright\arraybackslash}X@{}}
\toprule
Group & Quantity & Value \\
\midrule
Mechanical & $N,m,k,\gamma$ & $10,1,1,0.5$ \\
& Rayleigh coefficients $(a_R,b_R)$ & $(0.01,0.01)$ \\
& drive and sensor masses & 3 and 7 \\
& forcing amplitudes & $0.02,0.04,0.08,0.12$ \\
& interface and forcing band & $0.20$--$0.66$ rad/s \\
& model-order study band & $0.20$--$1.20$ rad/s \\
\addlinespace
ROM & basis and orders & mass-normalized modes; $r=1,2,3,4$ \\
& first four frequencies & $0.2846,0.5635,0.8308,1.0813$ rad/s \\
\addlinespace
Interface & $(\omega_{o,1},\omega_{o,2})$ & $(3.0,2.6)$ rad/s \\
& $(\zeta_{o,1},\zeta_{o,2})$ & $(0.50,0.40)$ \\
& $(g_1,g_2)$ and $\tau_{\rm sync}$ & $(4.0,3.2)$ and $1.5$ s \\
& distorted case $(\chi_1,\chi_2)$ & $(40,30)$ \\
& output channels & independently transduced $q_1,q_2$ \\
\bottomrule
\end{tabularx}
\end{table}

\section{Results}
\label{sec:results}

We organize the numerical results around three comparisons: the reduced and full mechanical source responses, predicted and reference dynamic readout outputs, and normalized-state distortion introduced by the interface. We first examine model order and basis dependence, then interface identification and mismatch, and finally the physical amplitude and phase changes associated with readout and forcing strength.

\subsection{Source-ROM convergence}
\label{sec:results_order}

\begin{figure}[H]
\centering
\begin{subfigure}{0.78\linewidth}
\centering
\includegraphics[width=\linewidth]{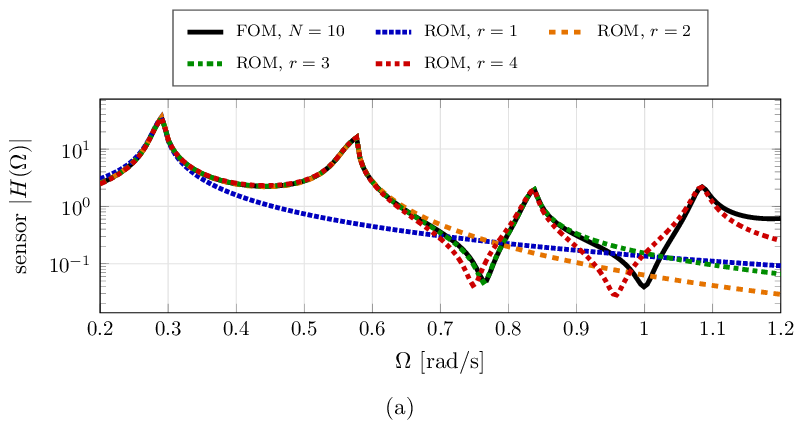}
\phantomcaption\label{fig:model_order_response}
\end{subfigure}
\par\medskip
\begin{subfigure}{0.78\linewidth}
\centering
\includegraphics[width=\linewidth]{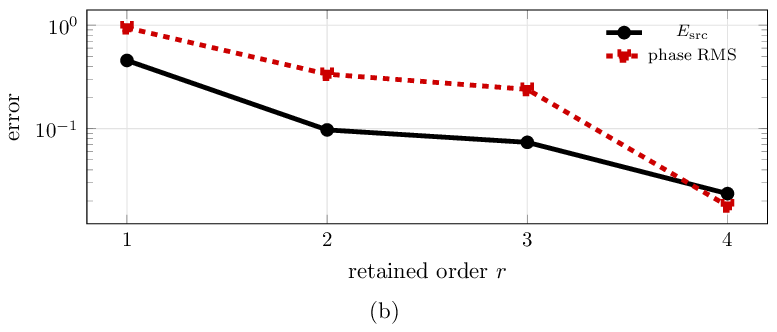}
\phantomcaption\label{fig:model_order_errors}
\end{subfigure}
\caption{Source-ROM order study at $F_0=0.04$ over $0.20$--$1.20$ rad/s. Panel~\ref{fig:model_order_response} shows full-order and reduced responses for $r=1,2,3,4$. Panel~\ref{fig:model_order_errors} shows source band error and phase RMS error.}
\label{fig:model_order}
\end{figure}

Figure~\ref{fig:model_order_response} compares the mass-7 displacement response of the full-order model with reduced responses retaining $r=1,2,3,4$ modes at $F_0=0.04$. Four linear natural frequencies lie within the $0.20$--$1.20$ rad/s band. The one-mode model represents the lowest-frequency response, and the additional modes improve representation of the remaining resonances. Differences from the full response include contributions from omitted modes and nonlinear interactions not resolved within each selected subspace.

Across this band, $E_{\rm src}$ is $45.6\%$, $9.73\%$, $7.37\%$, and $2.35\%$ for one, two, three, and four retained modes, respectively. The corresponding RMS phase errors are $0.943$, $0.336$, $0.240$, and $0.0179$ rad. Figure~\ref{fig:model_order_errors} shows how the reduced displacement prediction improves as the trial space captures more of the modes contributing to this four-resonance response. These source errors measure the complex physical observation, including contributions outside the two modal coordinates used for the normalized-state metric.

The mean normalized-state fidelities for two, three, and four retained modes are $\OrderTwoFidelity$, $0.99999996$, and $0.99999991$, respectively. They show close agreement in the normalized first-two-mode coefficient ratios for the evaluated ROMs. The larger physical source-response errors demonstrate that high two-component fidelity does not ensure an equally accurate displacement prediction when higher modes contribute to the observation. We therefore retain $E_{\rm src}$ and the modal-coefficient error $E_q$ as separate measures of reduced-source performance.

We benchmark runtime at the five frequencies $\{0.20,0.285,0.565,0.83,1.08\}$ rad/s. Each timing is the median of three uncached serial runs, including operator assembly, integration, and Fourier extraction. The one-, two-, three-, and four-mode models require $8.67$, $9.98$, $12.38$, and $12.62$ s, giving speedups of $1.53$, $\TwoModeSpeedup$, $1.07$, and $\FourModeSpeedup$ relative to the full model for this subset. These timings quantify the performance of the present ten-mass implementation and show that the more accurate higher-order ROMs provide only modest acceleration. The Supplementary Material reports the benchmark details.

The projected nonlinear-force evaluation retains all $N+1$ spring extensions, so reducing the number of coordinates does not remove the full-dimensional spring evaluation. This distinction between state reduction and nonlinear-force hyper-reduction is well established \cite{Chaturantabut2010}. We therefore interpret the timing results as implementation-specific rather than evidence of computational scaling to larger structures.

\subsection{Order selection across forcing and basis choice}
\label{sec:results_selection}

\begin{figure}[H]
\centering
\begin{subfigure}[b]{0.32\linewidth}
\centering
\includegraphics[width=\linewidth]{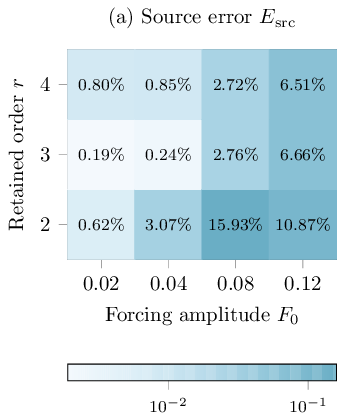}
\phantomcaption\label{fig:order_forcing_source}
\end{subfigure}\hfill
\begin{subfigure}[b]{0.32\linewidth}
\centering
\includegraphics[width=\linewidth]{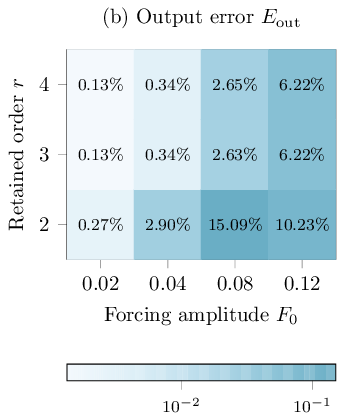}
\phantomcaption\label{fig:order_forcing_output}
\end{subfigure}\hfill
\begin{subfigure}[b]{0.32\linewidth}
\centering
\includegraphics[width=\linewidth]{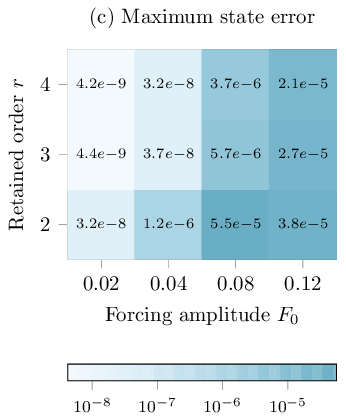}
\phantomcaption\label{fig:order_forcing_state}
\end{subfigure}
\caption{Dependence of source, output, and normalized-state errors on retained order and forcing amplitude over $0.20$--$0.66$ rad/s. Each cell uses independently converged full and reduced responses.}
\label{fig:order_forcing}
\end{figure}

Figures~\ref{fig:order_forcing_source}, \ref{fig:order_forcing_output}, and \ref{fig:order_forcing_state} compare the mass-7 source-response error, coupled readout-output error, and normalized first-two-mode state error, respectively, across forcing levels and retained modal orders. The source metric includes omitted spatial contributions to the physical observation, whereas the output metric reflects how the reference linear interface weights the first two modal coefficients. The state metric evaluates their normalized complex ratio. Using reference interface parameters isolates mechanical reduction effects in this comparison. We apply the prescribed offline tolerances in Section~\ref{sec:order_selection} and report the smallest accepted order in Table~\ref{tab:order_selection}.

\begin{table}[H]
\centering
\caption{Minimum tested modal order satisfying $E_{\rm src},E_{\rm out}\leq0.05$ and maximum ROM state error $\leq10^{-3}$.}
\label{tab:order_selection}
\begin{tabular}{@{}cc@{}}
\toprule
$F_0$ & minimum accepted $r$ \\
\midrule
\RomOrderSelectionRows
\bottomrule
\end{tabular}
\end{table}

On the sampled validation frequencies, two modal coordinates meet the selected criteria at $F_0=0.02$ and $0.04$, while three are required at $F_0=0.08$. At $F_0=0.12$, no tested modal order meets all response-error limits. The maximum normalized source-state error remains below $10^{-3}$ for every evaluated two-to-four-mode ROM, so state fidelity does not determine the accepted order in this numerical study. The three-mode ROM has a smaller narrow-band source error than the four-mode ROM at the two lower forcing levels, showing that error at a selected physical output need not decrease monotonically with retained dimension. At $F_0=0.12$, targeted quadrature refinement gives source errors of $\RefinedStrongTwoError$ and $\RefinedStrongFourError$ for two and four modes, both exceeding the tolerance. These conclusions are conditional on the sampled operating regime and its frequency resolution.

We construct an uncentered displacement POD basis from converged full-order snapshots at $F_0=0.02$ and $0.08$ and 15 training frequencies. We then hold the basis fixed and evaluate it at $F_0=0.04$ and $0.12$ using 75 frequencies excluded from training. This arrangement tests the fixed basis on the reported held-out set rather than establishing accuracy throughout the continuous band. We project the reconstructed POD displacement onto the same first two physical linear modes used for the state comparisons.

Table~\ref{tab:pod} compares modal and POD bases at equal retained dimension on the specified held-out conditions. Its first error pair gives the complex mass-7 source-response error $E_{\rm src}$, and the second gives the maximum sampled normalized-state error $1-F_\psi^{\rm ROM}$. These metrics respond to different discrepancies: physical displacement includes contributions from all participating modes, while the normalized-state measure evaluates the selected first-two-mode complex ratio. All reported comparisons use the same full-order references on each table's test frequencies.

At $F_0=0.04$, the modal basis yields smaller source-response errors for $r=2$ and $3$. For $r=4$, the POD source error is $0.465\%$, compared with $0.861\%$ for the modal basis. The modal basis gives smaller normalized-state errors at all three orders. We interpret these different rankings through the distinct comparisons represented by the metrics: a basis can improve mass-7 displacement prediction without producing the smallest error in the normalized first-two-mode state.

At $F_0=0.12$, both two-coordinate models give source-response errors near $12.8\%$ on the specified 75-frequency test set. The three-coordinate POD model gives $5.19\%$, and the four-coordinate POD model gives $2.05\%$, compared with $8.03\%$ for four linear modes. For the four-coordinate POD model, the reported readout-output error is $2.00\%$ and the maximum normalized-state error is $\PODStrongFourState$; it satisfies the illustrative tolerances on this test set. All models in Table~\ref{tab:pod} satisfy the normalized-state tolerance, so that criterion does not distinguish the models here. Because the source calculations exhibit strong-forcing sensitivity to frequency refinement elsewhere in the study, we interpret the POD advantage as a result for the reported held-out grid rather than a grid-independent strong-forcing guarantee.

\begin{table}[H]
\centering
\caption{Modal and POD bases at equal retained dimension on held-out frequencies and forcing levels. Sensor-response error $E_{\rm src}$ and maximum normalized-state error are evaluated relative to the full-order model.}
\label{tab:pod}
\small
\begin{tabular}{@{}cccccc@{}}
\toprule
& & \multicolumn{2}{c}{$E_{\rm src}$} & \multicolumn{2}{c}{maximum $1-F_\psi^{\rm ROM}$} \\
\cmidrule(lr){3-4}\cmidrule(l){5-6}
$F_0$ & $r$ & modal & POD & modal & POD \\
\midrule
\RomPodRows
\bottomrule
\end{tabular}
\end{table}

\subsection{Interface identification, noise, and model mismatch}
\label{sec:results_robustness}

For the matched noiseless interface at $F_0=0.04$, the held-out interface-only error is $E_{\rm out}=\InterfaceValidationError$ and the maximum relative error among seven identified parameters is $\ParameterRecoveryError$. These roundoff-level values verify the implementation under synthetic data generated by the same linear model used for fitting. When the fixed interface model is driven by the $r=2$ ROM, the held-out coupled-output error is $\CoupledValidationError$ over $0.20$--$0.66$ rad/s, compared with the source error $\InterfaceBandSourceError$. The remaining coupled discrepancy in this matched noiseless setting is associated with the reduced-source input.

We evaluate measurement-noise sensitivity using independent complex Gaussian perturbations with channelwise RMS levels of $0.1\%$, $1\%$, and $5\%$ and 30 trials per level. We fit parameters on calibration samples and evaluate the resulting outputs against clean held-out responses. At $1\%$ noise, the median output error is $\NoiseOneMedianError$, while the median maximum relative parameter error is $\NoiseOneParameterError$. These results distinguish predictive accuracy over the selected band from reliable recovery of individual physical parameters. The transducer resonances lie above the fitted band, and bounded optimization permits compensating parameter changes; the data do not establish a unique physical interpretation of every fitted parameter \cite{Ljung1999,NoelKerschen2017}.

We evaluate structural mismatch by fitting the specified bounded linear interface model to time-integrated cubic-interface responses at $F_0=0.04$ and holding the fitted parameters fixed at other forcing levels. The held-out error is $\MismatchBaseError$ at the calibration forcing and $\MismatchStrongError$ at $F_0=0.12$. The fitted solution reaches a parameter bound, so these errors characterize this constrained approximation rather than the best possible linear approximation without bounds. We use full-order source coefficients to isolate the discrepancy between the fitted linear readout response and the nonlinear reference interface.

\subsection{Interface-induced amplitude and phase distortion}
\label{sec:results_transfer}

\begin{figure}[H]
\centering
\begin{subfigure}{0.82\linewidth}
\centering
\includegraphics[width=\linewidth]{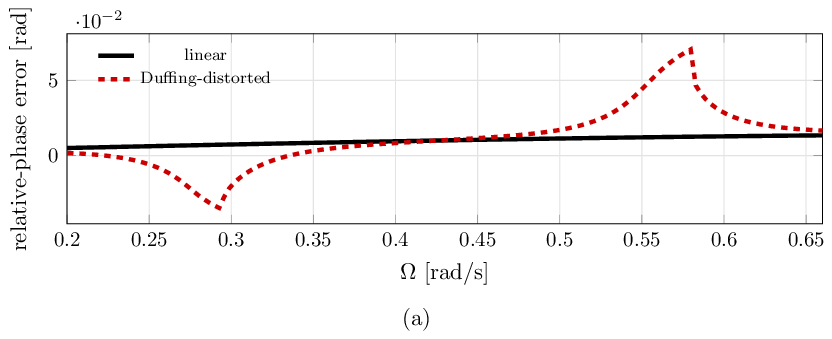}
\phantomcaption\label{fig:transfer_phase_residual}
\end{subfigure}
\par\medskip
\begin{subfigure}{0.82\linewidth}
\centering
\includegraphics[width=\linewidth]{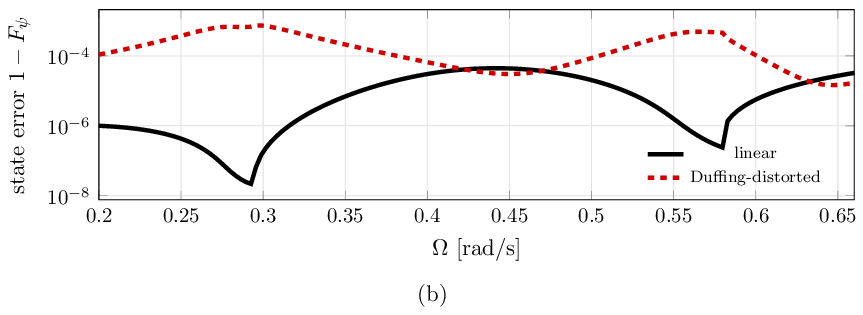}
\phantomcaption\label{fig:transfer_phase_state_error}
\end{subfigure}
\caption{Physical source-to-output transfer after static gain calibration and common-delay correction. Panel~\ref{fig:transfer_phase_residual} shows the relative-phase residual retaining the dynamic transducer phase. Panel~\ref{fig:transfer_phase_state_error} shows the state error $1-F_\psi$ on a logarithmic scale. Cubic interface distortion increases both errors.}
\label{fig:transfer_phase}
\end{figure}

We evaluate physical transfer after applying the static-gain normalization and common $1.5$~s delay correction in Eq.~\eqref{eq:calibrated_output_channel}. The remaining frequency-dependent dynamic phase is governed by Eq.~\eqref{eq:interface_transfer}. Unequal channel resonance and damping parameters produce a relative-phase residual across the source band. Because both fundamental coefficients are evaluated at the same drive frequency, an additional common delay does not change that relative phase.

At $F_0=0.04$, the linear interface gives $E_{\phi,{\rm MAE}}=\LinearPhaseMAE$ rad and $E_{\phi,{\rm RMS}}=\LinearPhaseRMS$ rad. With the cubic restoring term, these increase to $\DistortedPhaseMAE$ and $\DistortedPhaseRMS$ rad. In a first-harmonic approximation, the cubic term contributes an effective stiffness $3\chi_jA_j^2/4$ at fundamental output amplitude $A_j$ \cite{NayfehMook1979,Detroux2015}. We obtain the reported nonlinear outputs from direct time integration, including the influence of higher harmonics on the extracted fundamental. The phase metrics quantify only the fundamental two-channel state and do not measure complete-waveform harmonic distortion.

Figure~\ref{fig:transfer_phase_residual} shows the relative-phase residual of the linear and cubic interfaces. The linear case varies smoothly across the band, while the cubic case shows a negative excursion near the first resonance and a larger positive excursion near the second. The residual sign identifies the direction of the source-to-output phase change under the stated channel convention. These effects concern physical transduction even when the corresponding interface model predicts its output accurately.
Figure~\ref{fig:transfer_phase_state_error} shows the normalized-state error $1-F_\psi$. The cubic interface produces larger errors across most sampled frequencies. This error reflects the combined effect of relative amplitude and phase changes, and Eq.~\eqref{eq:channel_distortion} gives its linear-channel interpretation. We compute the nonlinear case directly from integrated source and output coefficients. State fidelity remains insensitive to common amplitude scaling, so it is interpreted alongside the amplitude-transfer metric $E_A$ and the phase results.

\subsection{Two-component state preservation}
\label{sec:results_state}

\begin{figure}[H]
\centering
\begin{subfigure}{0.86\linewidth}
\centering
\includegraphics[width=\linewidth]{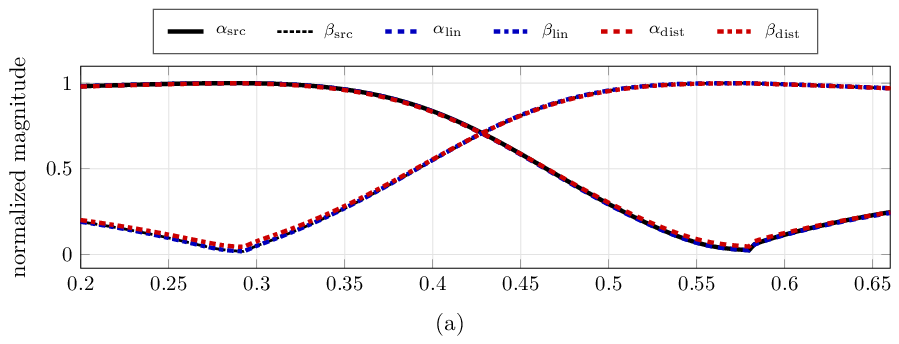}
\phantomcaption\label{fig:state_channels_magnitudes}
\end{subfigure}
\par\medskip
\begin{subfigure}{0.86\linewidth}
\centering
\includegraphics[width=\linewidth]{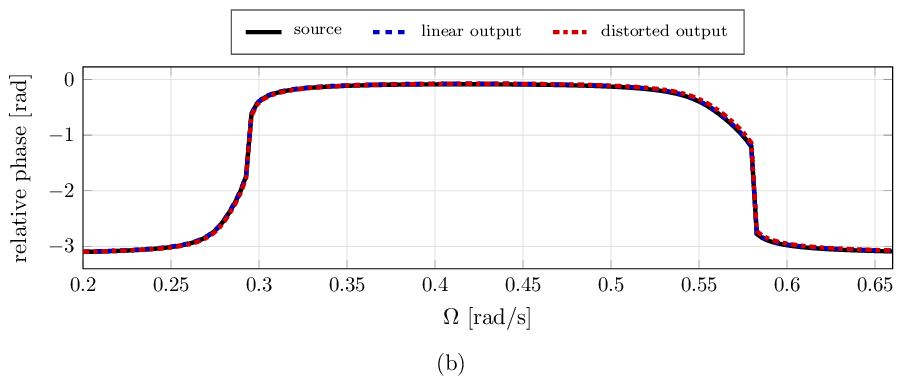}
\phantomcaption\label{fig:state_channels_phase}
\end{subfigure}
\caption{Normalized source and output channel magnitudes and relative phase across drive frequency. Static gain calibration, common-delay correction, and a shared normalization convention enable comparison of the states.}
\label{fig:state_channels}
\end{figure}

Figure~\ref{fig:state_channels_magnitudes} compares the normalized magnitudes $|\alpha|$ and $|\beta|$ for the source and the two readout cases. Their frequency dependence describes how the selected modal pair shares the normalized amplitude. The linear and nonlinear interfaces produce similar modal partitions over much of the band, with a visible additional imbalance for the cubic interface. The normalized plots concern amplitude ratios, not the absolute gain or attenuation of the readout signals.
Figure~\ref{fig:state_channels_phase} compares the corresponding source and output relative phases across drive frequency. The two readout cases follow the source variation, while the cubic interface introduces a larger systematic shift, consistent with Fig.~\ref{fig:transfer_phase_residual}. The reported mean fidelities are $\LinearMeanFidelity$ for the linear interface and $\DistortedMeanFidelity$ for the nonlinear interface. These high values describe agreement of the normalized modal pairs over the sampled operating points; they do not establish the same accuracy for absolute channel amplitudes or the entire displacement field. The curves form a frequency-dependent path $\psi(\Omega)$, with numerical convergence checks described in the Supplementary Material.

\subsection{Effects of source and interface nonlinearity}
\label{sec:results_amplitude}

\begin{figure}[H]
\centering
\begin{subfigure}{0.84\linewidth}
\centering
\includegraphics[width=\linewidth]{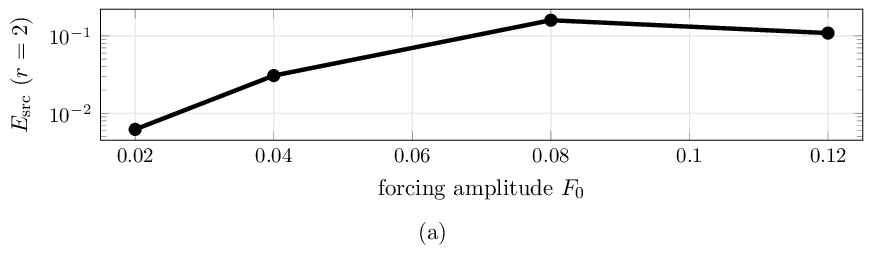}
\phantomcaption\label{fig:forcing_source_error}
\end{subfigure}
\par\medskip
\begin{subfigure}{0.84\linewidth}
\centering
\includegraphics[width=\linewidth]{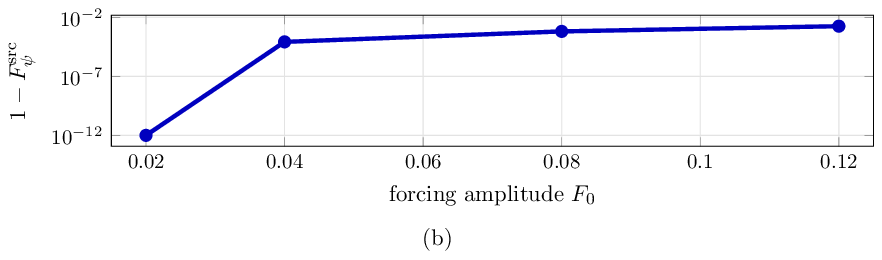}
\phantomcaption\label{fig:forcing_source_state}
\end{subfigure}
\par\medskip
\begin{subfigure}{0.84\linewidth}
\centering
\includegraphics[width=\linewidth]{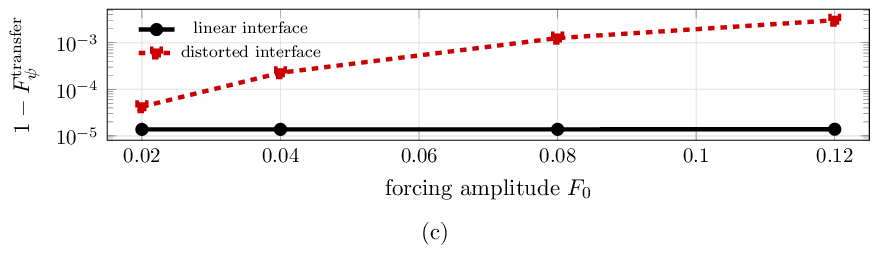}
\phantomcaption\label{fig:forcing_transfer_state}
\end{subfigure}
\caption{Forcing-dependent source prediction and two-channel state transfer over $0.20$--$0.66$ rad/s. Panel~\ref{fig:forcing_source_error} shows the two-mode source error on the plotted grid; targeted strong-forcing refinement is discussed in the text. Panel~\ref{fig:forcing_source_state} shows the sampled source-state deviation relative to the weak-forcing $F_0=0.02$ state path. Panel~\ref{fig:forcing_transfer_state} shows the sampled mean physical transfer distortion $1-F_\psi$ for linear and cubic readout interfaces. State-error panels use logarithmic axes.}
\label{fig:forcing}
\end{figure}

Figure~\ref{fig:forcing_source_error} shows the two-mode source-ROM error for the forcing sweep over $0.20$--$0.66$ rad/s. On the original common sample grid, the errors are $\WeakForceSourceError$ at $F_0=0.02$, $\HighForceSourceError$ at $F_0=0.08$, and $\StrongForceSourceError$ at $F_0=0.12$. Targeted refinement near strong nonlinear resonances changes the strongest-forcing result to $\RefinedStrongTwoError$ and reverses the apparent error decrease on the coarse grid. We therefore use the refined result when interpreting strong-forcing model adequacy and retain the coarse-grid value to document sampling sensitivity. Responses are independently initialized from rest at each frequency; possible additional steady-state branches are outside the validation protocol \cite{NayfehMook1979,Detroux2015,Kuether2015}.

Figure~\ref{fig:forcing_source_state} reports the mean normalized source-state error $1-F_\psi^{\rm src}$ relative to the $F_0=0.02$ reference path at the sampled frequencies. Its increase with forcing shows that the relative amplitude and phase of the selected two-mode state change across these operating conditions. This is a comparison between source states at different forcing levels, not an error of the readout interface.

Figure~\ref{fig:forcing_transfer_state} reports source-to-output normalized-state distortion at each forcing level. For the fixed linear interface, the sampled mean error changes little with forcing, consistent with the differential-transfer dependence in Eq.~\eqref{eq:channel_distortion}. For the cubic interface, the mean state error increases over the tested forcing levels, and its relative-phase RMS error rises from $\WeakForcePhase$ to $\StrongForcePhase$ rad. These results characterize the fundamental modal pair for the specified readout parameters.

\section{Concluding Remarks}
\label{conclusion}

We developed an offline output-based validation procedure for a nonlinear mechanical lattice coupled to a two-channel dynamic readout. The procedure evaluates full-order-referenced mechanical and readout prediction errors alongside physical changes in a normalized two-component modal state. The linear error decomposition distinguishes source truncation and interface-model contributions to the complex output discrepancy, while the channel-distortion identity relates normalized-state fidelity to unequal channel transfer factors. For the fundamental coefficients observed at a common frequency, common transport delay affects complex alignment but cancels from relative-phase and normalized-state comparisons.

The numerical comparisons show that ROM accuracy depends on the physical observation, retained basis, forcing amplitude, and validation grid. The four-mode source error is $\OrderFourError$ over the four-resonance band, and the two-mode source and coupled-output errors are $\InterfaceBandSourceError$ and $\CoupledValidationError$ over the two-resonance interface band. The modal-order acceptance map uses illustrative source and output tolerances; the normalized-state constraint does not determine the accepted order in the tested cases. The held-out POD comparison shows improved strong-forcing source accuracy for the four-coordinate POD basis on its specified test grid, while separate frequency-refinement results show that modal-ROM source errors are sensitive to frequency resolution near nonlinear resonances. At $F_0=0.04$, the nonlinear interface produces a larger mean absolute relative-phase error than the linear interface, while both retain high mean normalized-state fidelity. Runtime gains are modest for the present ten-mass implementation because nonlinear-force evaluation retains the full spring dependence.

The noise study shows that accurate held-out in-band output prediction can accompany substantial errors in individual interface parameters. These findings distinguish the predictive accuracy of the fitted interface on held-out frequencies within the calibration band from the accuracy of its estimated physical parameters. When we fit the specified bounded linear model to nonlinear interface data, the remaining held-out residual identifies mismatch within that constrained readout representation.

We conclude that source displacement error, modal-channel prediction error, readout-output error, and physical amplitude and phase transfer provide complementary evidence for evaluating phase-sensitive mechanical ROMs. The normalized two-component representation connects the simulated modal amplitude ratios and relative phases with established classical elastic-bit descriptions \cite{RN21,RN156}. The present calculations use directly specified simulated modal-channel inputs, fundamental steady-state coefficients, and independently initialized periodic responses. Further work can address reconstruction from sparse physical measurements, alternative nonlinear response branches, readout back-action, and larger systems. Within the tested configuration, the framework provides explicit and reproducible criteria for interpreting reduced-source accuracy together with dynamic readout behavior.

\newpage
\begingroup
\small

\section*{\small Author Contributions}

M.A.H. conceived the idea of the research. K.T.M performed the analytical study. All authors analyzed the findings and contributed to the scientific discussion and manuscript writing.

\section*{\small Competing Interests}

The authors declare no competing interests.

\section*{\small Data Availability}

The datasets used and analyzed during the current study available from the corresponding author on reasonable request.

\endgroup

\newpage

\bibliographystyle{unsrt}
\bibliography{references}

\end{document}